\documentclass{article}

\usepackage{arxiv}
\usepackage{natbib}
\usepackage[utf8]{inputenc} 
\usepackage[T1]{fontenc}    
\usepackage{hyperref}       
\usepackage{url}            
\usepackage{booktabs}       
\usepackage{amsfonts}       
\usepackage{nicefrac}       
\usepackage{microtype}      
\usepackage{lipsum}
\usepackage{microtype}
\usepackage{graphicx}
\usepackage{subfigure}
\usepackage{booktabs} 
\usepackage{url}
\usepackage{threeparttable}
\usepackage{multirow}
\usepackage{amsmath}
\usepackage{algorithm}
\usepackage{algorithmic}
\usepackage{bm}
\usepackage{array}
\usepackage{stfloats}
\usepackage{comment}
\usepackage{authblk}

\title{Adversarial Samples on Android Malware Detection Systems for IoT Systems}

\author[1]{Xiaolei Liu}
\author[2]{Xiaojiang Du}
\author[1]{Xiaosong Zhang}
\author[1]{Qingxin Zhu}
\author[3]{Mohsen Guizani}

\affil[1]{University of Electronic and Science Technology of China}
\affil[2]{Temple University, USA}
\affil[3]{University of Idaho, USA}

\begin{document}
\maketitle

\begin{abstract}
Many IoT(Internet of Things) systems run Android systems or Android-like systems. With the continuous development of machine learning algorithms, the learning-based Android malware detection system for IoT devices has gradually increased. However, these learning-based detection models are often vulnerable to adversarial samples. An automated testing framework is needed to help these learning-based malware detection systems for IoT devices perform security analysis. The current methods of generating adversarial samples mostly require training parameters of models and most of the methods are aimed at image data. To solve this problem, we propose a \textbf{t}esting framework for \textbf{l}earning-based \textbf{A}ndroid \textbf{m}alware \textbf{d}etection systems(TLAMD) for IoT Devices. The key challenge is how to construct a suitable fitness function to generate an effective adversarial sample without affecting the features of the application. By introducing genetic algorithms and some technical improvements, our test framework can generate adversarial samples for the IoT Android Application with a success rate of nearly 100\% and can perform black-box testing on the system.
\end{abstract}


\section{Introduction}
\label{sec:introduction}
Since many IoT(Internet of Things) devices run Android systems or Android-like systems, with the popularity of IoT devices, Android malware for IoT devices is also increasing. Meanwhile, machine learning has received extensive attention and has gained tremendous application development in many fields, such as financial economics, driverless, medical, and network security. So there are many learning-based Android malware detection systems \cite{Ham,McNeil,Aswini,Odusami1,misra2017unit,alhassan2018fuzzy}.

However, while machine learning brings us great convenience, it also exposes a lot of security problems \cite{Barreno2010}. Several papers have studied related Android and IoT security issues \cite{Wu2016,Cheng2017,Hei2013,Du2007,du2009transactions}. Scholars in the security field are increasingly concerned about the security issues associated with the lack of fairness and transparency in machine learning algorithms. An attacker can predict certain sensitive information by observing the model, or recover sensitive data in the data set through existing partial data. A current attack method is called a poisoning attack. Biggio B and Zhu attempted to attack the adaptive facial recognition system by poisoning attack \cite{Biggio2012,Biggio2013,Biggio2014,Xinzhong2013}. During the model update, they injected malicious data to offset the central value of the recognition feature in the model, so as to achieve the purpose of verifying the attacker's image. Biggio B and Nelson B also attacked the supervised learning algorithm SVM \cite{Biggio}. Experiments show that the test error of the model classifier can be significantly increased during the gradient rise. However, the injected sample data must meet certain constraints in order to deceive the model, and must be the attacker to control the label of the injection point. Yang Chaofei, Wu Qing et al. conducted an experiment on poisoning attacks against neural network learning algorithms\cite{Chaofei201703}. Compared with the direct gradient algorithm, this proposed method can increase the attack sample generation speed by about 240 times.

In fact, although a poisoning attack can make the model go wrong, the attacker has to work hard on how to inject malicious data. Another common method can let models get the wrong result in a short time, that is, adversarial sample attack. Christian Szegedy et al. first proposed the concept of adversarial samples \cite{Szegedy2013}. By deliberately adding minor changes in the dataset, the perturbed samples will cause the model to output a false result with high confidence. Adversarial samples can increase the prediction error of the model, so that the originally correctly classified sample migrates to the other side of the decision area, thereby being classified into another category.

Existing models are vulnerable to adversarial samples \cite{Carlini201608,Dezfooli2017,FARLEY2016,Papernot2016}. For example, in a malware recognition system, by adding a small perturbation to the original software sample, the result of the sample classification can be changed with a high probability, and even the sample can be classified into an arbitrarily designated label according to the attacker's idea. This makes adversarial samples attack a huge hazard to malware recognition systems \cite{chen2018automated,chen2018droideye,al2018adversarial}.

All the learning-based Android malware detection systems for IoT devices have the above problems, so a testing framework is needed to test the robustness of these detection systems. To address this challenge, we propose TLAMD, a \textbf{t}esting framework for \textbf{l}earning-based \textbf{A}ndroid \textbf{m}alware \textbf{d}etection systems for IoT Devices. When the test results show that the detection system cannot resist the attack of the adversarial samples, it indicates that this detection system has potential safety hazards, and it needs to be reinforced.

Therefore, how to generate effective adversarial samples is the core issue of the entire testing framework. Our approach to generating adversarial samples for the Android IoT malware detection model is based on genetic algorithm. Without the knowledge of the model parameters, the original sample is used as the input of the approach, and finally the adversarial sample of the specific label is generated. The information used is only the probability of the various types of labels output by the model. We hope that this method can be a robust benchmark for the learning-based Android malware detection model for IoT devices. Our contribution is mainly reflected as follows:

1. We migrated the application of adversarial samples from the image recognition domain to the Android malware detection domain of IoT devices. In this process, simply replacing the model's training data from a picture to an Android application is not possible. On the one hand, the data of the binary program is not continuous like the image data. On the other hand, random perturbation of the binary program may lead to the crash of the program, so special processing is required for the Android application to ensure the validity of the adversarial samples. We borrowed the processing method of Kathrin Grosse \cite{Grosse}, which realized the disturbance to the Android application by adding the request permission code in the $AndroidManifest.xml$ file. The difference is that we have made corresponding analysis and restrictions on the types and quantities of permissions that can be added. This method can ensure that the original function of the app is not affected and can be used normally; and the app can be disturbed in the simplest way to achieve the effect of changing the model detection result;

2. We introduce the genetic algorithm into the adversarial sample generation method and implement the black-box attack against the machine learning model. Without knowing the internal parameters such as the gradient and structure of the target network, it is only necessary to know the probability of various types of labels output by the model. Compared to Kathrin Grosse's approach, our approach not only implements black-box attacks, but also has a higher success rate, almost 100\%.

The rest of the paper is organized as follows. Section 2 introduces the related background of our approach. Section 3 presents TLAMD(A Testing Framework for Learning-based Android Malware Detection Systems for IoT Devices). Section 4 presents and discusses our experimental results. Finally, further discussions and conclusions are accomplished in Section 5.

\section{Related Background}
\label{sec:background}

\subsection{Neural Network}
The essence of the neural network is a function $y=F(x)$, the input $x$ is an n-dimensional vector, and the output $y$ is an m-dimensional vector. The function $F$ implies the model parameter $\theta$. The purpose of the training network is to calculate the value of the parameter $\theta$ from the known partial sample information. After the model is completed, the result of predicting $x$ is to solve the value of $y$ by the function $F$. In this paper, we mainly study the neural network of the $m$ classifier (that is, the output $y$ is an m-dimensional vector). The output of the last layer of the neural network uses a fully connected layer. The classifier outputs the index with the largest value in the output vector dimension as the result, that is:

\begin{equation}
	L(x) = arg \max_{j=1}[F(x)]_{i}
\end{equation}

where $L(x)$ is the category of $x$. 

Define $F$ as a single-layer fully-connected neural network. The output of the (n-1)-th layer is the input of the n-th layer, then:

\begin{equation}
	y_{n} = F_{n}(y_{n-1})
\end{equation}

Typical n-layer fully connected neural networks are:

\begin{equation}
	F = F_{n}*F_{n-1}*...*F_{2}*F_{1}
\end{equation}

\begin{equation}
	F_{n}(x) = \sigma(w_{n} * x + b_{n})
\end{equation}

where $\sigma$ is a linear or non-linear activation function. The commonly used activation functions are RELU\cite{MAAS2013}, tanh\cite{MISHKIN201511}, sigmoid, etc., $\omega$ is the weight of this layer, $b$ is the layer offset.

\subsection{Genetic Algorithm}

The idea of the genetic algorithm is to simulate the biological evolution process of natural selection. Using the thought of evolutionary theory, the process of finding the optimal solution of a certain objective function is simulated into the evolution process of the population. Based on the idea of the population, the algorithm uses a population containing information to perform an optimal search in multiple directions and completes the exchange and reconstruction of information in the search process.

The genetic algorithm can be used to search for the feasible solution space of a problem, and then find the possible optimal solution, which is the uncertainty optimization in the optimization problem. Uncertain optimization is to rely on random variables in the search direction, rather than a certain mathematical expression. Compared with other algorithms, the advantage is that when the optimization converges to the local extremum, the search result can jump out of an optimal solution and continue to search for a better feasible solution.

By choosing the appropriate objective function, the generation of the adversarial sample can be transformed into a solution to the optimization problem. The process of solving the optimal solution corresponding to the objective function is actually the process of generating the adversarial sample. This shows that genetic algorithms can be effectively applied to machine learning and other fields in terms of parameter optimization and function solving. In terms of parameter optimization, Zhi Chen, Tao Lin et al. used a parallel genetic algorithm to optimize the parameter selection of Support Vector Machine (SVM) \cite{Chen2016}. Experiments show that the proposed method is superior to the grid search in classification accuracy, the number of selected features and running time. Aaron Vose, Jacob Balma et al. optimized the hyperparameters of the neural network in deep learning based on genetic algorithms. Anh Viet Phan et al. proposed a GA-SVM model which can effectively improve classification performance based on genetic algorithm and SVM classifier \cite{Viet2017}. Francisco Villegas Alejandre, Nareli Cruz Cortes et al. selected features based on machine learning to detect botnets\cite{Alejandre2017}. A genetic algorithm is used in this method to select the set of features that provide the highest detection rate.

\subsection{Adversarial Samples}

On many machine learning models, the decision boundary of the classifier has a certain margin of error. That is, when the disturbance satisfies $||\eta||_\infty < \epsilon$, the classifier considers that the perturbed input $x' = x + \eta$ is the same as the original input $x$. Therefore, when the perturbation value on each feature element is less than $\epsilon$, the classifier cannot discern the difference in the sample. However, changes in input characteristics have a cumulative effect on model predictions. Although the perturbation value on each feature element is small, the accumulated error is sufficient to influence the model prediction result.

On each neuron, the adversarial sample will have the following operations:

\begin{equation}
	\omega^{T} x' = \omega^{T}(x+\eta)
\end{equation}

Although the adversarial sample has no effect on the classification results of the single-dimensional neuron classifier. However, deep learning has a considerable number of neurons. The weight in each neuron has $n$ dimensions. If the average variation of an element in the weight vector is $m$, the activation effect will increase by $n*m$. And in a high dimensional linear classifier, each individual input feature is normalized. The result is that in the process of deep learning, a small change may not be enough to change the input result, but multiple disturbances to the input will cause the classifier to make a wrong classification result.
 
Many methods of generating adversarial samples need to know the parameters of the learning model to calculate the perturbation values, but some subsequent studies have shown that without knowing the parameters of the learning model\cite{Papernot2017,Papernot201602,Papernot201605,Narodytska2017}. The attacker can interact with the black-box learning model to calculate the samples. Specifically, the attacker can estimate the boundary of the decision region of the model according to the difference of the model output brought by different samples, and then use the estimated boundary as a substitute model. Finally, the adversarial samples are calculated by the parameters of the substitute model. Considering that more and more malicious Android application detection methods based on machine learning, how to evaluate the robustness of these detection methods becomes a new problem. Since most machine learning algorithms are vulnerable to adversarial samples, we have thought of using the generated adversarial samples to test the robustness of these detection methods.

\section{Methodology}

\subsection{Framework}

\begin{figure}[hbt]
\centering
  \includegraphics[scale=0.85]{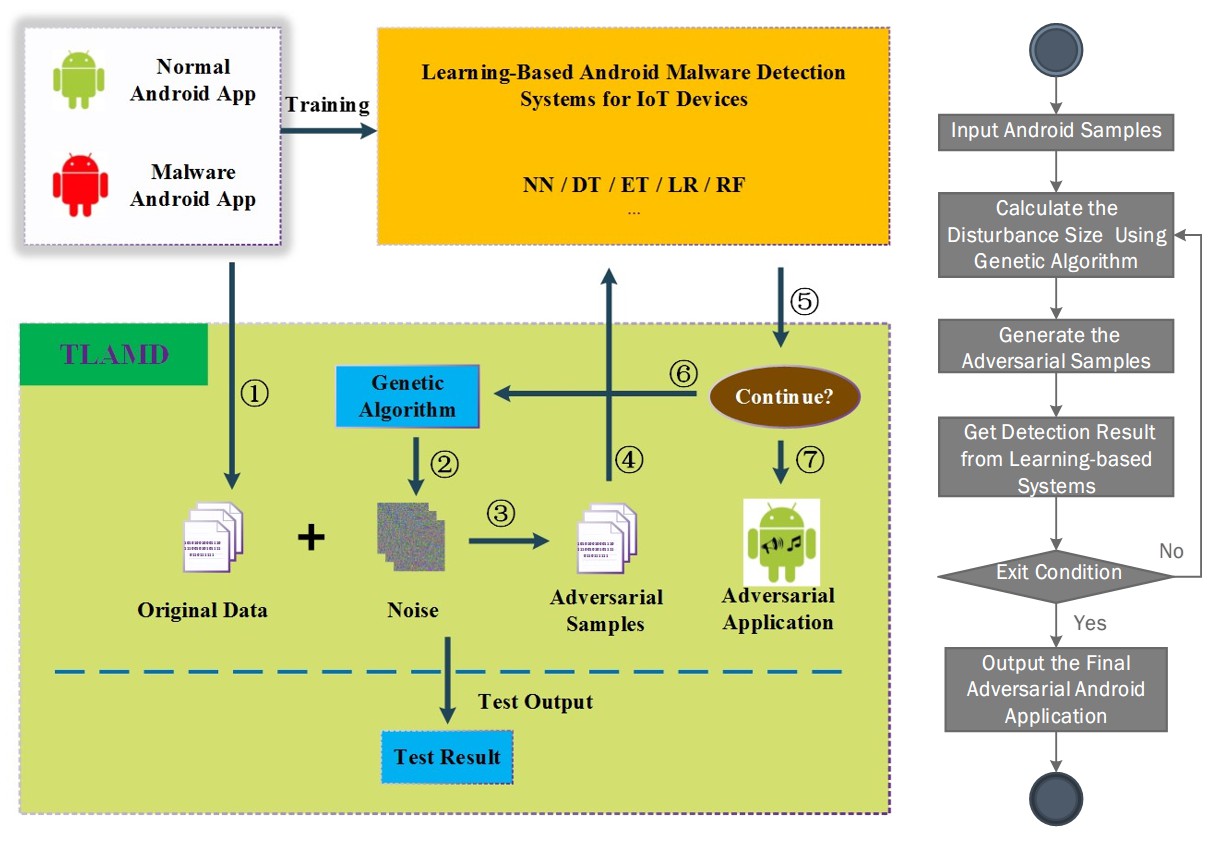}
  \caption{Overview of our testing framework for learning-based Android Malware detection systems for IoT devices. 1) Original Sample Input; 2) Calculate the disturbance size; 3) Generate the adversarial samples; 4) Get detection result from learning-based systems; 5) Determine if the exit condition is met; 6) If not, calculate the new disturbance size using genetic algorithm; 7) If yes, output the final adversarial android application.}
  \label{p6}
\end{figure}

The overview of TLAMD is shown in Figure\ref{p6}. 

 When the test results show that the detection system cannot resist the attack against the adversarial sample, it indicates that the system has potential safety hazards and it is necessary to implement such reinforcement measures as distillation defense\cite{PAPERNOT} on the detection system. As we can see, how to generate an adversarial sample is the main challenge of this testing framework. Therefore, we will describe the algorithm in detail for generating an adversarial sample for Android malware.

\subsection{Algorithm}

Our goal is to add minor perturbations to the malware without changing the malware functionality, so that the previously trained detection model misidentifies it as normal software. Therefore, our approach generates an adversarial sample by adding permission features to the $AndroidManifest.xml$, and in order not to affect the function of the original malware, the disturbance does not reduce the existing permission features. For a single input sample $X$, the classifier returns a two-dimensional vector $F(X)=[F_{0}(X), F_{1}(X)]$, where $F_{0}( X)$ indicates the probability that the software is a normal software, $F_{1}(X)$ indicates the probability that the software is a malware, and satisfies the constraint $F_{0}(X)+F_{1}(X)= 1$. We aim to add a perturbation $\delta$ to make the classification result  $F_{1}(X+\delta)$ is less than $F_{0} (X+ \delta)$. At the same time, the smaller the $\delta$, the better, that is, the fewer the number of permission features added in the manifest file, the better. For example, for a specific malware $x$, we use genetic algorithm to find out which permission features $\delta$ are added to $x$, and finally make $x$ detected as normal software with minimum number of permission added.

From a mathematical point of view, the process of misjudging the detection model by adding the permission features is regarded as a problem to be solved. The feasible solution space of the problem is the disturbance if the detection model is successfully misjudged. The optimal solution is to minimize the disturbance value, that is, add the least permission feature. A genetic algorithm is a type of algorithm that finds the possible optimal solution by searching for a feasible solution space of a problem. Our approach is to use genetic algorithms to search for the minimum perturbation value that causes the detection model to be misjudged. The pseudo code of our approach is shown in Algorithm\ref{alg1}.

\begin{algorithm}
    \caption{Generating an adversarial sample.}
    \label{alg1}
    \begin{algorithmic}
        \REQUIRE Popluation Size $pop\_size$
        \STATE $\delta \gets initialization()$
        \FOR{$i = 0 \to pop\_size$ }
		\STATE $P_i \gets Crossover\_Operator()$ \
		\STATE $P_i \gets Mutation\_Operator()$\
		\STATE \textbf{Compute}$\ \to S(\delta)$

        \IF{$F(X+\delta) > 1 - F(X+\delta)$}
            \STATE \textbf{Continue}
        \ELSE
            \STATE \textbf{Output}$\ \to \delta$
        \ENDIF
        
        \ENDFOR
    \end{algorithmic}
\end{algorithm}

The specific steps are as follows:

1) Randomly generate the population $\delta=P_1, P_2,..., P_M$. $M$ is the number of individuals, the individual $P_{i} \in\{0,1\}^{n}$ refers to the permission characteristics to be added in the category, and $n$ is the number of permission features in the category. 1 means to add the corresponding permission, otherwise 0 means not to add. Our strategy is to only add permissions and not reduce permissions. Therefore, if the original malicious sample has a certain permission feature, the permission cannot be removed, that is, the disturbance is 0.

2) Determine the fitness function.

\begin{equation}
	S(\delta) = \min\ {w_{1} \cdot F(X+\delta) + w_{2} \cdot num(\delta)}
\end{equation}

where $w_{1}$ and $w_{2}$ represent the two weights, $\delta$ is the added small disturbance, $F(X+\delta) \in [0,1]$ means that the probability of original malicious sample is still detected as a malware, $num(\delta_{i})$ indicates the number of permission features added.

When $w_1$ is much larger than $w_2$, the sample after the addition of the disturbance must be detected as normal by the detection model to survive, and the individual detected as a malicious sample will be eliminated. The surviving individual must meet the minimum number of added permission features, otherwise it will also be eliminated. The fitness function defined in this way searches for an optimal solution that can successfully cause the detection model to be misjudged. 

3) Perform mutation operations according to a certain probability to generate new individuals. The mutation refers to adding a disturbance to the corresponding category according to a certain probability, that is, changing the value from 0 to 1, and satisfying the constraint proposed in step 1).

4) Generate a new generation of the population from the mutation and return to step 2). If the preset number of iterations is reached, the loop is exited.

\section{Experiments}

\subsection{Data Set and Environment}

In order to verify the effectiveness of the adversarial sample, we attempt to train five different classifier models, including logistic regression (LR), decision tree (DT), and fully connected neural network (NN) and so on. The hardware environment and software environment of all experiments are as follows:

\begin{table}[h]
\caption{The environment of all experiments.}
\fontsize{8}{12}\selectfont 
\centering
\begin{tabular}{c|c}
\textbf{CPU}	& Inter(R) Core(TM) i5-7400 CPU @ 3.00GHz\\
\textbf{Memery}	& 8GB\\
\textbf{Video Card}	& Inter(R) HD Graphics 630\\
\textbf{Operating System}	& Windows 10\\
\textbf{Programming Language}	& Python 3.6\\
\textbf{Development Platform}	& Jupyter Notebook\\
\textbf{Dependence}	& Tensorflow, Keras, numpy etc.\\
\end{tabular}
\label{environment_table}
\end{table}

All the data we use in the experiments come from the DREBIN dataset\cite{Arp2014,Spreitzenbarth2013}. The DREBIN dataset has a total of 123,453 sample data for Android applications, including 5,560 malicious samples and contains as many as 545,333 behavioral features.

The features of the Android app in this dataset consist of eight categories:

S1) Hardware components, which are used to set the hardware permissions required by the software.

S2) Requested permissions, which are granted by the user at the time of installation and allow the application software to access the corresponding resources.

S3) App components, which include four different types of interfaces: activities, services, content providers and broadcast receivers.

S4) Filtered intents, which are used for process communication between different components and applications.

S5) Restricted API calls, access to a series of key API calls.

S6) Used permissions, a subset of permissions that are actually used and requested in S5.

S7) Suspicious API calls, API calls for allowing access to sensitive data and resources.

S8) Network addresses, the IP addresses accessed by the application, including the hostname and URL.

The first four classes are extracted from the manifest file, and the last four classes are extracted from the disassembly code. Since our method only adds permission requests to the $AndroidManifest.xml$ file, we only cover the features in S1 to S4. In Section\ref{feature_extraction}, we further reduce the feature categories used.

\begin{table}[hbt]
\caption{8 features in DREBIN dataset}
\fontsize{8}{12}\selectfont 
\setlength{\tabcolsep}{3pt}
\center
  \begin{tabular}{|c|c|c|c|}
  \hline
   Class & Name & Numbers & Rate(/Total) \\
    \hline
   S1 & Hardware Components & 72 & 0.013\% \\
   S2 & Requested Permissions & 3812 & 0.704\% \\
   S3 & App Components & 218951 & 40.488\% \\
   S4 & Filtered Intents & 6379 & 1.178\% \\
   S5 & Restricted API Calls & 733 & 0.136\% \\
   S6 & Used Permissions & 70 & 0.013\% \\
   S7 & Suspicious API Calls & 315 & 0.058\% \\
   S8 & Network Address & 310447 & 57.4\%\\
\hline
  \end{tabular}
\end{table}

\subsection{Android Malware Detection Model}

First, a detection model is trained to determine whether an Android sample is malware. When the detection model reaches a certain accuracy, our approach is used to generate an adversarial sample for the model.

\subsubsection{Feature Extraction}
\label{feature_extraction}

We use a random forest approach to measure the importance of features in the feature extraction phase. The number of features is effectively reduced without affecting the accuracy of detection.

Random forest is an integrated learning in machine learning. It is an integrated classifier composed of multiple sets of decision trees: ${h(X,\theta _k), k=1,2,...}$, where $\theta _k$ is a random variable subject to independent and identical distribution, and $k$ represents the number of decision trees. The principle is to generate multiple decision trees and let them learn independently and make corresponding predictions. Finally, observe which category is selected the most and get the result.

The specific steps are as follows:

1) Select out of bag (OOB) to calculate the corresponding out-of-bag data deviation $error _1$ for each decision tree.

2) Add random noise, perturb all samples of OOB, and then calculate the out-of-bag data deviation $error_2$ again.

3) Define and calculate the importance of the features: 

\begin{equation}
	I = \sum(error_1 - error_2)/N
\end{equation}

where $N$ is the number of forest decision trees.

If $error_2$ is greatly increased after adding random noise, the OOB accuracy rate decreases, indicating that this type of feature has a greater impact on the prediction result, that is, the importance is higher.

\begin{figure}[hbt]
\center
  \includegraphics[width=0.8\textwidth]{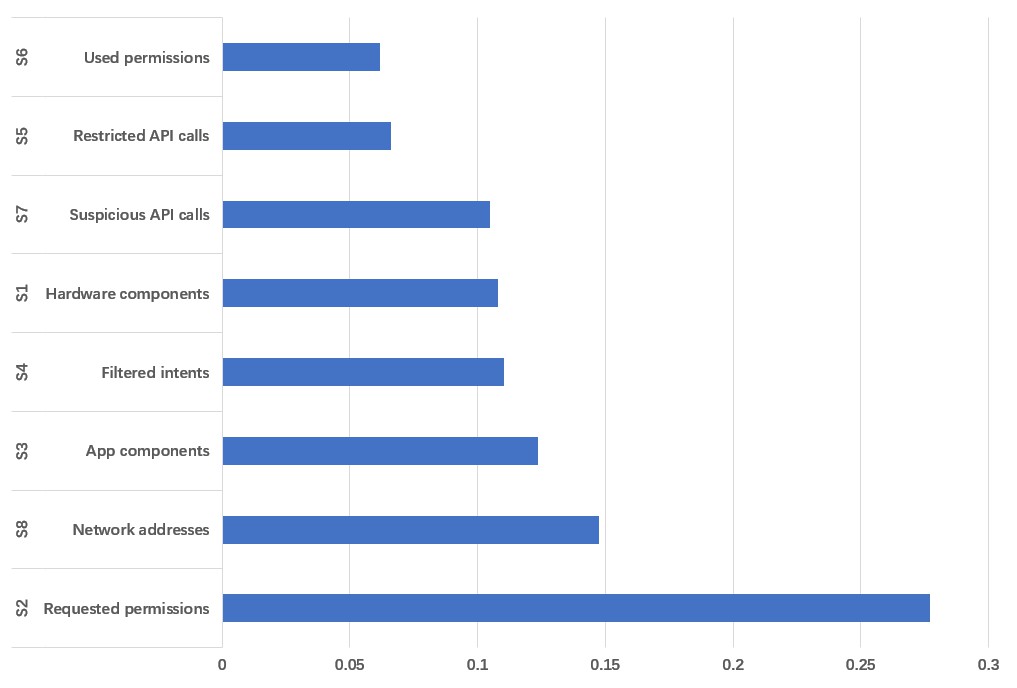}
  \caption{The sorting result of feature importance.
  The ordinate represents different behavioral feature categories and the abscissa represents the proportion of importance.}
  \label{p2}
\end{figure}

The sorting result of feature importance is shown in Figure\ref{p2}.

As we mentioned before, we only cover the four types of features from S1 to S4. Taking into account the number and importance of various features, we finally choose the two characteristics of S1 and S2.

\subsubsection{Training Detection Model}

To test the effectiveness of our method for different detection models, we trained five kinds of detection models.

a. Neural Network

Our neural network chooses a two-layer fully connected model with 200 neurons in each connected layer and the activation function is RELU. The output layer of the last layer has two neurons and is the $soft\max$ activation unit. And no dropout operation is performed on each layer. To train our network, we used the gradient descent training method with a batches size of 256. All data was trained 5 times per iteration.

b. Logistic Regression

Since there are only two types of target predictions, we adopt a two-class logistic regression model. The penalty term selects the L2 paradigm, and the model parameters satisfy the Gaussian distribution, that is, the parameters are constrained so that they do not over-fitting. Considering that the solution problem is not a linear multi-core, and the number of samples is selected to be larger than the number of features, the dual method is not set. Set the condition for stopping the solution is that the loss function is less than or equal to $1 \times e^{-4}$; the category weight defaults to 1. The maximum number of iterations of the algorithm convergence is set to 10.

c. Decision Tree

The decision tree is a tree structure used for classification. The maximum depth of the decision tree is set to 15 to prevent overfitting. The $min\_impurity\_decrease$ is set to 0. The $min\_samples\_split$ is set to 2, indicating the minimum number of samples required for internal node subdivision. The $min\_samples\_leaf$ is set to 10, indicating the minimum number of samples in the leaf node. The $max\_leaf\_nodes$ is set to None, which is expressed as the maximum number of leaf nodes in the decision tree. The $min\_weight\_fraction\_leaf$ is set to 0, which represents the minimum value of all sample weights and sums of leaf nodes.

d. Random Forest

Random forest is an integrated learning. Through the bootstrap resampling technique, a number of sample inputs are randomly selected from the original training set with repeated iterations. In this way, a new training set is obtained, and then several decision trees are generated to form a random forest. The $max\_feature$ is set to auto, that is, a single decision tree can utilize all permission features. The $n\_estimators$ is set to 20, which means there are 20 decision trees to form the random forest to be trained. The $min\_sample\_leaf$ is set to 20, that is, the minimum number of sample leaves in each decision tree is 20.

e. Extreme Tree

Extra Tree is equivalent to a variant of the random forest. Compared with random forests, the randomness is further calculated when dividing the local best, that is, the selection of the division points is calculated. Most of its parameters are the same as those of random forests, except that $n\_estimators$ is set to 10 and $max\_depth$ is set to 50.

Finally, when the five detection models are trained, we test 42570 samples and the results are shown in Table\ref{test_results_table}:

\begin{table}[hbt]
\caption{The detection results of five models.}
\fontsize{8}{12}\selectfont  
\label{test_results_table}
\center
\setlength{\tabcolsep}{2.6mm}{
  \begin{tabular}{|c|cccc|ccc|}
  \hline
   Models$^{\mathrm{a}}$  & TP & FP & FN & TN & Accuracy & Precision & Recall \\
    \hline
   NN  & 40770 & 0 & 74 & 1726 & 99.83\% & 1 & 95.95\%\\
    \hline
  LR  & 40770 & 0 & 234 & 1566 & 99.45\% & 1 & 96.32\%\\
   \hline
  DT  & 40770 & 0 & 60 & 1740 & 99.86\% & 1 & 95.91\%\\
   \hline
  RF  & 40770 & 0 & 32 & 1768 & 99.92\% & 1 & 95.85\%\\
   \hline
  ET  & 40770 & 0 & 16 & 1784 & 99.96\% & 1 & 95.81\%\\
   \hline
   \multicolumn{6}{p{235pt}}{$^{\mathrm{a}}$NN = Neural Network, LR = Logistic Regression, DT = Decision Tree, RF = Random Forest, ET = Extreme Tree. }
  \end{tabular}}
\end{table}

\subsection{Simulation Experiments}

After getting the trained detection models, we will generate adversarial samples for the five models. The  features we add to the $AndroidManifest.xml$ file are from S1 or S2. The parameters of the generation algorithm are also different depending on the permission category. The details are as shown in Table\ref{parameters_table}.

\begin{table}[hbt]
\caption{The parameters of our approach.}
\fontsize{8}{12}\selectfont   
\label{parameters_table}
\setlength{\tabcolsep}{3pt}
\center
  \begin{tabular}{ccc}
  \hline
   Features  & S1: Hardware components & S2: Requested permissions \\
    \hline
  Initialize Probability & 1\% & 0.01\%  \\
  Mutation Probability  & 30\% & 0.5\% \\
  Iterations  & 50 & 50 \\
  Population  & 150 & 150 \\
   Attacked Samples  & 1000 & 1000 \\
   \hline
  \end{tabular}
\end{table}

The final experimental results are shown in Table\ref{results_table}. In the ten sets of adversarial sample generation experiments for the five detection models, the success rates are above 80\%, and most of them are close to 100\%. In order to generate these adversarial samples, the average number of permission features added is less than three. On the one hand, it shows that the adversarial sample generated by our method is very effective and our approach is able to be a robust benchmark for the learning-based Android malware detection model for IoT devices; on the other hand, it shows that the existing machine learning algorithms are very vulnerable to the adversarial sample. Our TLAMD test framework is very necessary. 

In subsequent experiments, we also performed a reinforcement method for the distillation defense of these models. However, the reinforced model is still unable to resist the attack of adversarial samples, and the success rate of our approach is still close to 100\%. This means that when we want to reinforce existing machine learning models, common methods such as distillation defenses work poorly. We need to find a more effective defense method.

\begin{table}[hbt]
\caption{The results of our approach.$^{\mathrm{a}}$}
\fontsize{8}{12}\selectfont 
\label{results_table}
\setlength{\tabcolsep}{3pt}
\center
  \begin{tabular}{|c|c|c|c|}
  \hline
 Model  & Category & Success Rate & Average of $num(\delta)$ \\
    \hline
NN  & S1 & 1 & 2.25  \\
  &S2  &1 & 2.33 \\
  \hline
LR  &S1  & 0.998 & 2.66 \\
  &S2  & 0.995 & 1.94 \\
  \hline
 DT  &S1  & 0.896 & 1.05 \\
   &S2  & 0.992 & 1.68 \\
   \hline
 RF  &S1  & 0.866 & 2.89 \\
   &S2  & 0.995 & 9.54 \\
   \hline
  ET  &S1  & 0.833 & 2.81 \\
   &S2  & 0.945 & 9.36 \\
   \hline
   \multicolumn{4}{p{200pt}}{$^{\mathrm{a}}$Each line of data in the table is the average of the 1000 sample tests results. }
  \end{tabular}
\end{table}

\begin{figure}[hbt]
\center
  \includegraphics[width=0.8\textwidth]{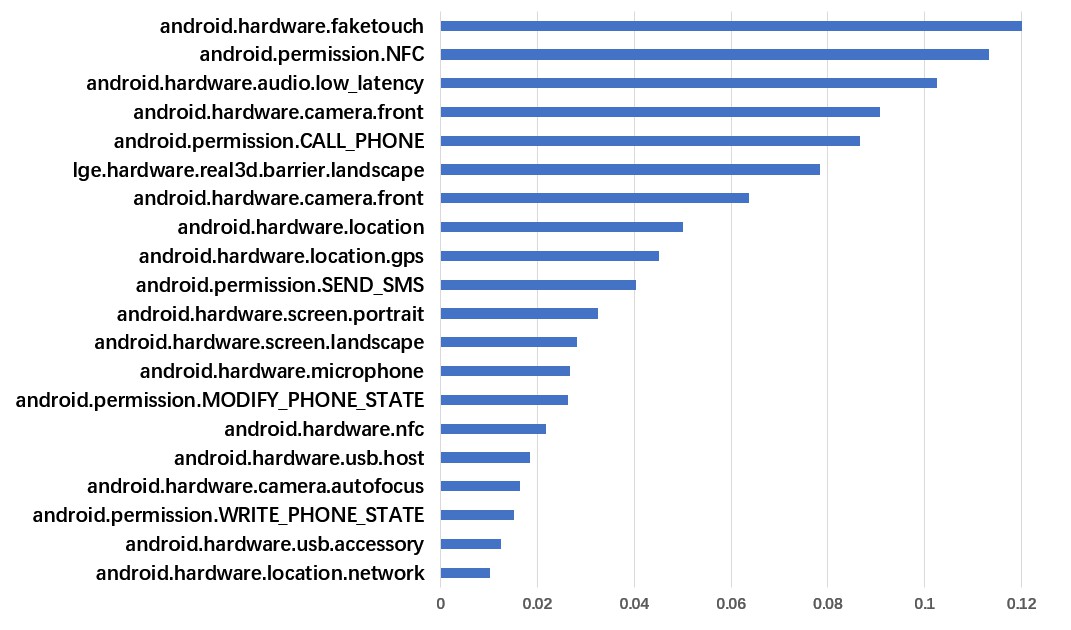}
  \caption{The most frequently added permissions in our adversarial sample generation experiments. The data is the average of $5 \times$ 2 $\times$ 1000 samples test results.}
  \label{p10}
\end{figure}

Fig.\ref{p10} shows the most frequently added permissions in the ten sets of adversarial sample generation experiments for the five kinds of detection models. Compared to other permission features, these permissions are mostly permissions that involve sensitive privacy. In order to verify whether these features have a decisive influence on the model discrimination results, we have conducted further experiments. In the new experiment, we will not allow the algorithm to add the features listed in the figure. However, the success rate of the generated adversarial samples is consistent with the previous one in Table\ref{results_table}, and the number of permission features added is slightly increased. It can be seen that those features that are added more frequently only have greater weight, but have no decisive influence on the results.

\begin{figure}[hbt]
\center
  \includegraphics[width=0.8\textwidth]{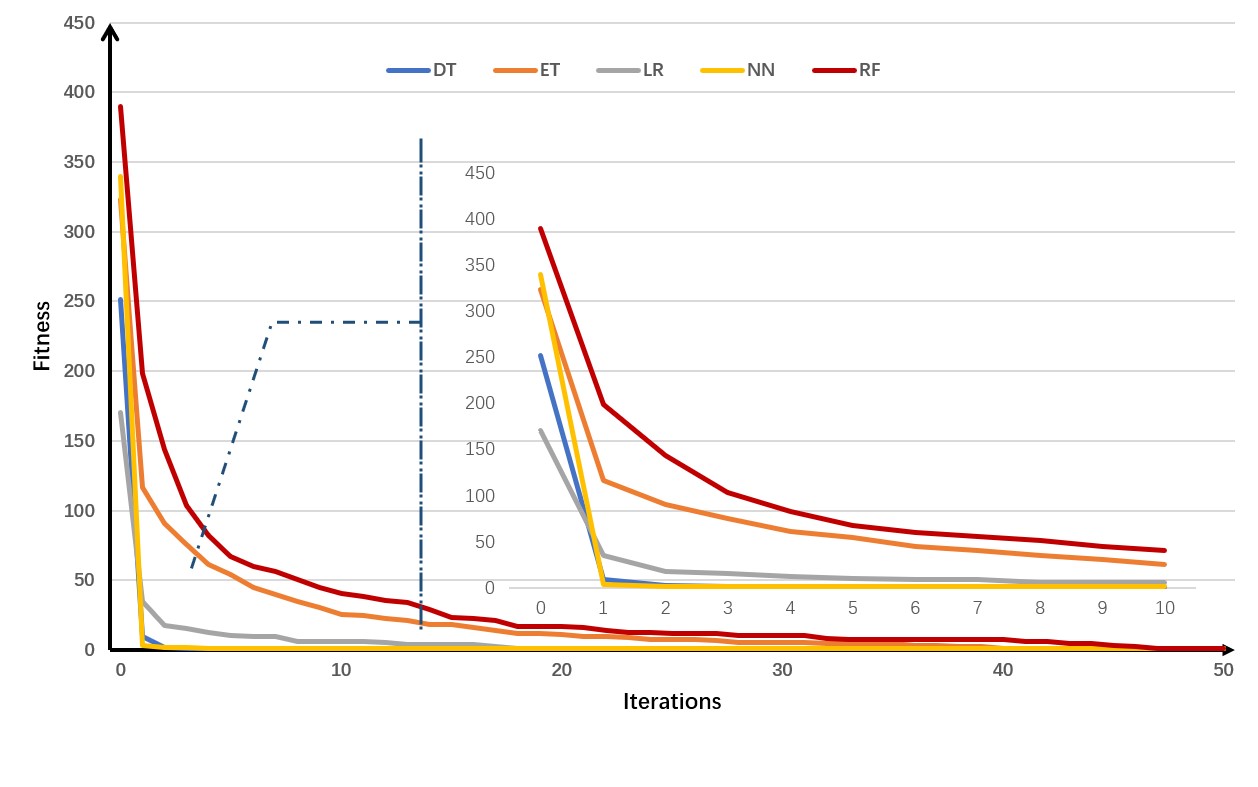}
  \caption{Trend graph of fitness function values with number of iterations.}
  \label{p11}
\end{figure}

Fig.\ref{p11} is a trend graph of fitness function values as a function of the number of iterations. As the number of iterations increases, the value of the fitness function decreases rapidly. It shows that it is very effective to use the genetic algorithm to solve the problem of generating adversarial samples.

By comparing the individual models, it can be found that the more complex the detection model, the better the effect of the adversarial samples generated for the model. This phenomenon may be different from what we expected. We believe that one possible reason is that the more complex the model, the more times the feature is processed. This makes small changes in features easily magnified, making the model very sensitive to adversarial samples.

\begin{figure}[hbt]
\center
  \includegraphics[width=0.8\textwidth]{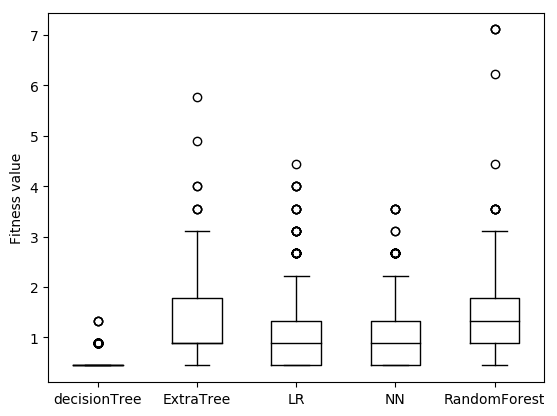}
  \caption{The fitness function values of adversarial samples for five detection models with S1 permission features. }
  \label{p7}
\end{figure}

\begin{figure}[hbt]
\center
  \includegraphics[width=0.8\textwidth]{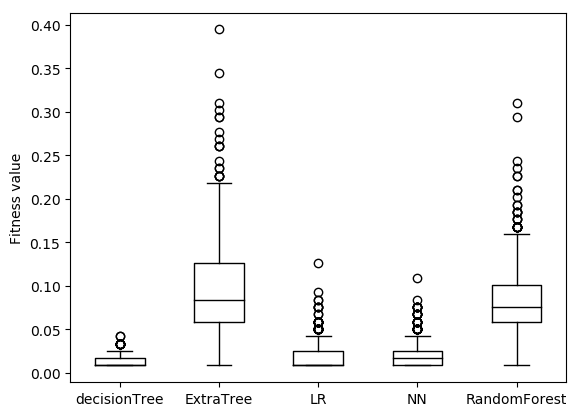}
  \caption{The fitness function values of adversarial samples for five detection models with S2 permission features. }
  \label{p9}
\end{figure}

Fig.\ref{p7} is a box plot of the fitness function values of adversarial samples for five detection models with S1 permission features and Fig.\ref{p9} is with S2 permission features. As can be seen from the figures, the adversarial samples generated by our approach is very stable.  There are only a very small number of divergence points out of 1000 samples. By comparing Fig.\ref{p7} and Fig.\ref{p9}, the stability of the adversarial sample generated by S2 is better. The reason is that the number of permission features in the S2 list is much larger than the number in the S1 list. This is equivalent to finding the optimal solution of the objective function in a larger space, so there is a greater probability of finding a better solution. Combined with Fig.\ref{p10}, it also provides us with an idea of how to strengthen the learning-based detection model. It is not useful to improve the defense of high-weight permission features. It is necessary to optimize the detection model so that it is not sensitive to small disturbances of all sample features.

\section{Conclusions}

To address the challenge of the lack of the testing framework for learning-based Android malware detection systems for IoT devices, we approach TLAMD. Our experimental results show that our approach generates high-quality adversarial samples with a success rate of nearly 100\% by adding permission features. In the technical implementation of the TLAMD algorithm, the selection of feature and the range of disturbance are the keys to have a good result. We hope TLAMD can be a benchmark for learning-based IoT Android malware detection model. The limitation of TLAMD is our black-box approach need frequent model requests and our future work includes reducing the requesting times and designing an effective defense approach to reinforce the malware detection model.

\bibliographystyle{authordate1}

\bibliography{mybibfile}

\begin{thebibliography}{}

\bibitem[\protect\citename{Al-Dujaili {\em et~al.\ }\relax,
  }2018]{al2018adversarial}
Al-Dujaili, Abdullah, Huang, Alex, Hemberg, Erik, \& O’Reilly, Una-May. 2018.
\newblock Adversarial deep learning for robust detection of binary encoded
  malware.
\newblock {\em Pages  76--82 of:} {\em 2018 IEEE Security and Privacy Workshops
  (SPW)}.
\newblock IEEE.

\bibitem[\protect\citename{Alejandre {\em et~al.\ }\relax,
  }2017]{Alejandre2017}
Alejandre, Francisco~Villegas, Cort{\'e}s, Nareli~Cruz, \& Anaya,
  Eleazar~Aguirre. 2017.
\newblock Feature selection to detect botnets using machine learning
  algorithms.
\newblock {\em Pages  1--7 of:} {\em Electronics, Communications and Computers
  (CONIELECOMP), 2017 International Conference on}.
\newblock IEEE.

\bibitem[\protect\citename{Alhassan {\em et~al.\ }\relax,
  }2018]{alhassan2018fuzzy}
Alhassan, JK, Misra, Sanjay, Umar, A, Maskeli{\=u}nas, Rytis,
  Dama{\v{s}}evi{\v{c}}ius, Robertas, \& Adewumi, Adewole. 2018.
\newblock A Fuzzy Classifier-Based Penetration Testing for Web Applications.
\newblock {\em Pages  95--104 of:} {\em International Conference on Information
  Theoretic Security}.
\newblock Springer.

\bibitem[\protect\citename{Arp {\em et~al.\ }\relax, }2014]{Arp2014}
Arp, Daniel, Spreitzenbarth, Michael, Hubner, Malte, Gascon, Hugo, Rieck,
  Konrad, \& Siemens, CERT. 2014.
\newblock DREBIN: Effective and Explainable Detection of Android Malware in
  Your Pocket.
\newblock {\em Pages  23--26 of:} {\em Ndss},  vol. 14.

\bibitem[\protect\citename{Aswini \& Vinod, }2015]{Aswini}
Aswini, AM, \& Vinod, P. 2015.
\newblock Towards the Detection of Android Malware using Ensemble Features.
\newblock {\em Journal of Information Assurance \& Security}, {\bf 10}(1).

\bibitem[\protect\citename{Barreno {\em et~al.\ }\relax, }2010]{Barreno2010}
Barreno, Marco, Nelson, Blaine, Joseph, Anthony~D, \& Tygar, J~Doug. 2010.
\newblock The security of machine learning.
\newblock {\em Machine Learning}, {\bf 81}(2), 121--148.

\bibitem[\protect\citename{Biggio {\em et~al.\ }\relax, }2012a]{Biggio2012}
Biggio, Battista, Fumera, Giorgio, Roli, Fabio, \& Didaci, Luca. 2012a.
\newblock Poisoning adaptive biometric systems.
\newblock {\em Pages  417--425 of:} {\em Joint IAPR International Workshops on
  Statistical Techniques in Pattern Recognition (SPR) and Structural and
  Syntactic Pattern Recognition (SSPR)}.
\newblock Springer.

\bibitem[\protect\citename{Biggio {\em et~al.\ }\relax, }2012b]{Biggio}
Biggio, Battista, Nelson, Blaine, \& Laskov, Pavel. 2012b.
\newblock Poisoning attacks against support vector machines.
\newblock {\em arXiv preprint arXiv:1206.6389}.

\bibitem[\protect\citename{Biggio {\em et~al.\ }\relax, }2013]{Biggio2013}
Biggio, Battista, Didaci, Luca, Fumera, Giorgio, \& Roli, Fabio. 2013.
\newblock Poisoning attacks to compromise face templates.
\newblock {\em Pages  1--7 of:} {\em Biometrics (ICB), 2013 International
  Conference on}.
\newblock IEEE.

\bibitem[\protect\citename{Biggio {\em et~al.\ }\relax, }2014]{Biggio2014}
Biggio, Battista, Fumera, Giorgio, \& Roli, Fabio. 2014.
\newblock Pattern recognition systems under attack: Design issues and research
  challenges.
\newblock {\em International Journal of Pattern Recognition and Artificial
  Intelligence}, {\bf 28}(07), 1460002.

\bibitem[\protect\citename{Carlini \& Wagner, }2017]{Carlini201608}
Carlini, Nicholas, \& Wagner, David. 2017.
\newblock Towards evaluating the robustness of neural networks.
\newblock {\em Pages  39--57 of:} {\em 2017 IEEE Symposium on Security and
  Privacy (SP)}.
\newblock IEEE.

\bibitem[\protect\citename{Chen {\em et~al.\ }\relax, }2018a]{chen2018droideye}
Chen, Lingwei, Hou, Shifu, Ye, Yanfang, \& Xu, Shouhuai. 2018a.
\newblock Droideye: Fortifying security of learning-based classifier against
  adversarial android malware attacks.
\newblock {\em Pages  782--789 of:} {\em 2018 IEEE/ACM International Conference
  on Advances in Social Networks Analysis and Mining (ASONAM)}.
\newblock IEEE.

\bibitem[\protect\citename{Chen {\em et~al.\ }\relax,
  }2018b]{chen2018automated}
Chen, Sen, Xue, Minhui, Fan, Lingling, Hao, Shuang, Xu, Lihua, Zhu, Haojin, \&
  Li, Bo. 2018b.
\newblock Automated poisoning attacks and defenses in malware detection
  systems: An adversarial machine learning approach.
\newblock {\em computers \& security}, {\bf 73}, 326--344.

\bibitem[\protect\citename{Chen {\em et~al.\ }\relax, }2016]{Chen2016}
Chen, Zhi, Lin, Tao, Tang, Ningjiu, \& Xia, Xin. 2016.
\newblock A parallel genetic algorithm based feature selection and parameter
  optimization for support vector machine.
\newblock {\em Scientific Programming}, {\bf 2016}.

\bibitem[\protect\citename{Cheng {\em et~al.\ }\relax, }2017]{Cheng2017}
Cheng, Yingxin, Fu, Xiao, Du, Xiaojiang, Luo, Bin, \& Guizani, Mohsen. 2017.
\newblock A lightweight live memory forensic approach based on hardware
  virtualization.
\newblock {\em Information Sciences}, {\bf 379}, 23--41.

\bibitem[\protect\citename{Du {\em et~al.\ }\relax, }2007]{Du2007}
Du, Xiaojiang, Xiao, Yang, Guizani, Mohsen, \& Chen, Hsiao-Hwa. 2007.
\newblock An effective key management scheme for heterogeneous sensor networks.
\newblock {\em Ad Hoc Networks}, {\bf 5}(1), 24--34.

\bibitem[\protect\citename{Du {\em et~al.\ }\relax, }2009]{du2009transactions}
Du, Xiaojiang, Guizani, Mohsen, Xiao, Yang, \& Chen, Hsiao-Hwa. 2009.
\newblock Transactions papers a routing-driven Elliptic Curve Cryptography
  based key management scheme for Heterogeneous Sensor Networks.
\newblock {\em IEEE Transactions on Wireless Communications}, {\bf 8}(3),
  1223--1229.

\bibitem[\protect\citename{Grosse {\em et~al.\ }\relax, }2017]{Grosse}
Grosse, Kathrin, Papernot, Nicolas, Manoharan, Praveen, Backes, Michael, \&
  McDaniel, Patrick. 2017.
\newblock Adversarial examples for malware detection.
\newblock {\em Pages  62--79 of:} {\em European Symposium on Research in
  Computer Security}.
\newblock Springer.

\bibitem[\protect\citename{Ham {\em et~al.\ }\relax, }2014]{Ham}
Ham, Hyo-Sik, Kim, Hwan-Hee, Kim, Myung-Sup, \& Choi, Mi-Jung. 2014.
\newblock Linear SVM-based android malware detection for reliable IoT services.
\newblock {\em Journal of Applied Mathematics}, {\bf 2014}.

\bibitem[\protect\citename{Hei {\em et~al.\ }\relax, }2013]{Hei2013}
Hei, Xiali, Du, Xiaojiang, Lin, Shan, \& Lee, Insup. 2013.
\newblock PIPAC: Patient infusion pattern based access control scheme for
  wireless insulin pump system.
\newblock {\em Pages  3030--3038 of:} {\em INFOCOM}.

\bibitem[\protect\citename{Maas {\em et~al.\ }\relax, }2013]{MAAS2013}
Maas, Andrew~L, Hannun, Awni~Y, \& Ng, Andrew~Y. 2013.
\newblock Rectifier nonlinearities improve neural network acoustic models.
\newblock {\em Page ~3 of:} {\em Proc. icml},  vol. 30.

\bibitem[\protect\citename{McNeil {\em et~al.\ }\relax, }2016]{McNeil}
McNeil, Paul, Shetty, Sachin, Guntu, Divya, \& Barve, Gauree. 2016.
\newblock SCREDENT: Scalable Real-time Anomalies Detection and Notification of
  Targeted Malware in Mobile Devices.
\newblock {\em Procedia Computer Science}, {\bf 83}, 1219--1225.

\bibitem[\protect\citename{Mishkin \& Matas, }2015]{MISHKIN201511}
Mishkin, Dmytro, \& Matas, Jiri. 2015.
\newblock All you need is a good init.
\newblock {\em arXiv preprint arXiv:1511.06422}.

\bibitem[\protect\citename{Misra {\em et~al.\ }\relax, }2017]{misra2017unit}
Misra, Sanjay, Adewumi, Adewole, Maskeli{\=u}nas, Rytis,
  Dama{\v{s}}evi{\v{c}}ius, Robertas, \& Cafer, Ferid. 2017.
\newblock Unit Testing in Global Software Development Environment.
\newblock {\em Pages  309--317 of:} {\em International Conference on Recent
  Developments in Science, Engineering and Technology}.
\newblock Springer.

\bibitem[\protect\citename{Moosavi-Dezfooli {\em et~al.\ }\relax,
  }2017]{Dezfooli2017}
Moosavi-Dezfooli, Seyed-Mohsen, Fawzi, Alhussein, Fawzi, Omar, \& Frossard,
  Pascal. 2017.
\newblock Universal adversarial perturbations.
\newblock {\em arXiv preprint}.

\bibitem[\protect\citename{Narodytska \& Kasiviswanathan,
  }2017]{Narodytska2017}
Narodytska, Nina, \& Kasiviswanathan, Shiva~Prasad. 2017.
\newblock Simple Black-Box Adversarial Attacks on Deep Neural Networks.
\newblock {\em Pages  1310--1318 of:} {\em CVPR Workshops}.

\bibitem[\protect\citename{Odusami {\em et~al.\ }\relax, }2018]{Odusami1}
Odusami, Modupe, Abayomi-Alli, Olusola, Misra, Sanjay, Shobayo, Olamilekan,
  Damasevicius, Robertas, \& Maskeliunas, Rytis. 2018.
\newblock Android Malware Detection: A Survey.
\newblock {\em Pages  255--266 of:} Florez, Hector, Diaz, Cesar, \&
  Chavarriaga, Jaime (eds), {\em Applied Informatics}.
\newblock Cham: Springer International Publishing.

\bibitem[\protect\citename{Papernot {\em et~al.\ }\relax, }2016a]{PAPERNOT}
Papernot, Nicolas, McDaniel, Patrick, Wu, Xi, Jha, Somesh, \& Swami, Ananthram.
  2016a.
\newblock Distillation as a defense to adversarial perturbations against deep
  neural networks.
\newblock {\em Pages  582--597 of:} {\em 2016 IEEE Symposium on Security and
  Privacy (SP)}.
\newblock IEEE.

\bibitem[\protect\citename{Papernot {\em et~al.\ }\relax, }2016b]{Papernot2016}
Papernot, Nicolas, McDaniel, Patrick, Jha, Somesh, Fredrikson, Matt, Celik,
  Z~Berkay, \& Swami, Ananthram. 2016b.
\newblock The limitations of deep learning in adversarial settings.
\newblock {\em Pages  372--387 of:} {\em Security and Privacy (EuroS\&P), 2016
  IEEE European Symposium on}.
\newblock IEEE.

\bibitem[\protect\citename{Papernot {\em et~al.\ }\relax,
  }2016c]{Papernot201602}
Papernot, Nicolas, McDaniel, Patrick, Goodfellow, Ian, Jha, Somesh, Celik,
  Z~Berkay, \& Swami, Ananthram. 2016c.
\newblock Practical black-box attacks against deep learning systems using
  adversarial examples.
\newblock {\em arXiv preprint}.

\bibitem[\protect\citename{Papernot {\em et~al.\ }\relax,
  }2016d]{Papernot201605}
Papernot, Nicolas, McDaniel, Patrick, \& Goodfellow, Ian. 2016d.
\newblock Transferability in machine learning: from phenomena to black-box
  attacks using adversarial samples.
\newblock {\em arXiv preprint arXiv:1605.07277}.

\bibitem[\protect\citename{Papernot {\em et~al.\ }\relax, }2017]{Papernot2017}
Papernot, Nicolas, McDaniel, Patrick, Goodfellow, Ian, Jha, Somesh, Celik,
  Z~Berkay, \& Swami, Ananthram. 2017.
\newblock Practical black-box attacks against machine learning.
\newblock {\em Pages  506--519 of:} {\em Proceedings of the 2017 ACM on Asia
  Conference on Computer and Communications Security}.
\newblock ACM.

\bibitem[\protect\citename{Phan {\em et~al.\ }\relax, }2017]{Viet2017}
Phan, Anh~Viet, Le~Nguyen, Minh, \& Bui, Lam~Thu. 2017.
\newblock Feature weighting and SVM parameters optimization based on genetic
  algorithms for classification problems.
\newblock {\em Applied Intelligence}, {\bf 46}(2), 455--469.

\bibitem[\protect\citename{Spreitzenbarth {\em et~al.\ }\relax,
  }2013]{Spreitzenbarth2013}
Spreitzenbarth, Michael, Freiling, Felix, Echtler, Florian, Schreck, Thomas, \&
  Hoffmann, Johannes. 2013.
\newblock Mobile-sandbox: having a deeper look into android applications.
\newblock {\em Pages  1808--1815 of:} {\em Proceedings of the 28th Annual ACM
  Symposium on Applied Computing}.
\newblock ACM.

\bibitem[\protect\citename{Szegedy {\em et~al.\ }\relax, }2013]{Szegedy2013}
Szegedy, Christian, Zaremba, Wojciech, Sutskever, Ilya, Bruna, Joan, Erhan,
  Dumitru, Goodfellow, Ian, \& Fergus, Rob. 2013.
\newblock Intriguing properties of neural networks.
\newblock {\em arXiv preprint arXiv:1312.6199}.

\bibitem[\protect\citename{Warde-Farley \& Goodfellow, }2016]{FARLEY2016}
Warde-Farley, David, \& Goodfellow, Ian. 2016.
\newblock 11 adversarial perturbations of deep neural networks.
\newblock {\em Perturbations, Optimization, and Statistics},  311.

\bibitem[\protect\citename{Wu {\em et~al.\ }\relax, }2016]{Wu2016}
Wu, Longfei, Du, Xiaojiang, \& Wu, Jie. 2016.
\newblock Effective defense schemes for phishing attacks on mobile computing
  platforms.
\newblock {\em IEEE Transactions on Vehicular Technology}, {\bf 65}(8),
  6678--6691.

\bibitem[\protect\citename{Yang {\em et~al.\ }\relax, }2017]{Chaofei201703}
Yang, Chaofei, Wu, Qing, Li, Hai, \& Chen, Yiran. 2017.
\newblock Generative poisoning attack method against neural networks.
\newblock {\em arXiv preprint arXiv:1703.01340}.

\bibitem[\protect\citename{Zhu, }2013]{Xinzhong2013}
Zhu, Xinzhong. 2013.
\newblock Super-class Discriminant Analysis: A novel solution for
  heteroscedasticity.
\newblock {\em Pattern Recognition Letters}, {\bf 34}(5), 545--551.

\end{thebibliography}

\end{document}